\documentclass[aps,prb,twocolumn,showpacs,superscriptaddress]{revtex4}
\usepackage{epsfig}
\usepackage{graphicx}
\usepackage{amsmath}
\usepackage{amsfonts}
\usepackage{amssymb}
\usepackage{bm}
\usepackage{bbold}

\begin{document}
\title{Constructing Landau framework for topological order: Quantum chains and ladders}
\author{Gennady Y. Chitov}
\affiliation{Department of Physics, Laurentian University, Sudbury, Ontario,
P3E 2C6 Canada}
\author{Toplal Pandey}
\affiliation{Department of Physics, Laurentian University, Sudbury, Ontario,
P3E 2C6 Canada}
\date{\today}
%
%
\begin{abstract}
We studied quantum phase transitions in the antiferromagnetic dimerized spin-$\frac12$ $XY$ chain and
two-leg ladders. From analysis of several spin models we
present our main result: the framework to deal with topological orders and hidden symmetries within the
Landau paradigm. After mapping of the spin Hamiltonians onto the tight-binding models
with Dirac or Majorana fermions and, when necessary, the mean-field approximation, the analysis can be done
analytically. By utilizing duality transformations the calculation of nonlocal string order parameters is mapped onto the
local order problem in some dual representation and done without further approximations. Calculated phase diagrams,
phase boundaries, order parameters and their symmetries for each of the phases provide a comprehensive quantitative Landau
description of the quantum critical properties of the models considered. Complementarily, the phases with hidden
orders can also be distinguished by the Pontryagin (winding) numbers which we have calculated as well.
This unified framework can be straightforwardly applied for various spin chains and ladders, topological insulators
and superconductors. Applications to other systems are under way.
\end{abstract}
\maketitle
%
%
\section{Introduction}
%
%
According to the Landau theory of phase transitions, the phases on both sides
of a critical point can be distinguished by different types of the long-range
order, or its absence. The ordering is described by an appropriately chosen local
order parameter. In the Landau framework phases are distinct due to their
different symmetries, and a continuous phase transition is always related to
spontaneous breaking of system's Hamiltonian symmetry. \cite{Landau5}

However, there is a mounting number of examples of systems where the states of
matter and their change cannot be classified in terms of the apparent symmetry
breaking and/or local order parameter, most notably a particular class of
gapped systems known as quantum spin liquids.\cite{Wen,FradkinBook13} One of
the best known example of such quantum liquids is realized in the Fractional
Quantum Hall Effect. Most likely, such a quantum (resonating valence
bond) spin liquid is also seen in the pseudogap phase of the high-$T_c$
cuprates. \cite{Wen} Low-dimensional and/or frustrated systems provide a lot of
possibilities for realization of various exotic quantum liquid states,
where a local long-range order is prevented even at zero temperature. In
low-dimensional quantum magnets, fermions, and topological insulators/superconductors
the existence of gapped phases and transitions between them are not necessarily
accompanied by breaking of some symmetry associated with local observables. See
\cite{Wen,FradkinBook13,Dagotto,Laflorencie2011,LaddersQC,SOPladders,Kim,Delgado,2Dspins,Kitaev06,Xiang07,TI}
and more references there.

To characterize these states a concept of \textit{topological order}
was introduced. \cite{Wen}  The exact definition of this concept is a really subtle issue.
\cite{Nussinov09} It is often associated with the nontrivial
Pontryagin, Chern or similar topological numbers (indices). \cite{FradkinBook13,TI,VolovikBook}
Hatsugai and co-workers \cite{Hatsugai-1} proposed the quantized Berry phase as a quantum order parameter
to probe nonlocal orders. This scheme was successfully applied, e.g., for several quantum spin liquid systems.
It turns out that the change of the quantized
Berry phase can diagnose also crossovers between the states which are not
separated by gap closing, local symmetry or order changes.
\cite{Hatsugai-2,Ezawa13,Mila13}

It has been also suggested that
topological entanglement, derived from the von Neumann entropy, provides another
quantitative measure of topological order since quantum critical points
separating different phases can be identified with extrema of the
entanglement entropy. \cite{Entang}

To characterize phases with hidden nonlocal or topological orders
the concept of string order parameter (SOP) introduced by
den Nijs and Rommelse is particularly instrumental. \cite{denNijs89}
The appearance of such nonlocal order is accompanied by a hidden symmetry breaking.
\cite{HiddenSSB} SOPs are successfully applied to characterize quantum phases
with hidden orders in various spin chains and ladders at zero
temperature.\cite{SOPladders,Laflorencie2011,Kim,Delgado,Xiang07,HiddenSSB,Hida92,GYC11}

The important property of SOP is that it can be used for both quantum and thermal phase transitions.
For instance, the SOP can be used for analysis of the thermal transitions into a Kosterlitz-Thouless-type
phase with an algebraic order.\cite{denNijs89} Very importantly, the SOP is a proper order parameter in the
sense of Landau: for the integrable models one can calculate exactly the
critical indices $\beta$ and $\eta$ from the SOP and the string-string correlation
function, resp., near the critical point \cite{Hida92} and
check explicitly that such $\beta$, $\eta$ along with other critical indices satisfy
the standard hyperscaling relations. \cite{GYC11} Since the SOP is a limit of a
nonlocal correlation function, its analytical calculation is a difficult task
even for exactly solvable models. For example, for the dimerized Heisenberg
spin-$\frac12$ chain the calculation is quite simple in the leading sine-Gordon
approximation for the Hamiltonian \cite{Hida92}, while taking into account the
marginal terms renders the problem much more difficult \cite{Bortz07}, and more
work is still needed. The other subtlety is that only for a spin chain or
a two-leg ladder the definition of the SOPs is straightforward, while already
for a three-leg ladder \cite{Kim,Delgado,GYC11}, to say nothing about a 2D lattice,
there are many ways to define the SOP. Potential multitude of SOPs and their
critical properties need further analysis.

There is an extensive recent literature on hidden orders in massive phases of
Heisenberg ladders.  In particular, the critical properties of the dimerized
two- and three-leg ladders have been studied in numerical and analytical works.
\cite{Delgado,GYC11,Kotov99,Cabra99,Nersesyan00,Okamoto03,Nakamura03,GYC08,MoreDimLadd}
Quantum critical lines or even gapless regions in the parameter space adjacent to
various massive phases of spin ladders may be also induced by frustrations,
magnetic fields, or four-spin interactions. \cite{Laflorencie2011,LaddersQC}

The nonlocal SOP can be defined for bosonic systems as well. \cite{Berg08,Rath13}
Moreover, this nonlocal order was even observed for bosons in optical lattice.
\cite{Enders11}

Another interesting class of models where nonlocal string-like order parameters
turned out to be essential are the probabilistic cellular automata which have been
experiencing a steadily growing interest during the last several decades.
It was shown  recently \cite{UsPerc} that the active phases of the (1+1)-dimensional automation models
possess numerous percolative nonlocal order parameters emerging as cascades of geometric phase transitions.

There has been an opinion expressed in the literature that various
low-dimensional fermionic or spin systems  with hiddenly ordered phases
and transitions between them, defy completely the Landau paradigm. It appears to
us that such a claim is too radical, and one can formulate the theory of
such transitions consistent with the Landau framework using the nonlocal SOPs.\cite{ChenHu07}
The Landau picture needs however to be amended, since some facets of
topological order quantified by topological indices, Berry phases,
entanglement, are not reducible even to nonlocal SOPs. The major problem is
that the current treatment of such systems in most cases is heavily numerical
and the qualitative physical picture is obscured. It is really warranted to put
more weight first on advancing relatively simple analytical approaches
based on the effective (mean-field) models, similar, e.g.,  to the Kitaev model
\cite{Kitaev06,Xiang07}, or the single-particle tight-binding models used
for the analyses of the topological insulators and superconductors.\cite{TI}
Results based on those models clearly indicate that interactions are not an indispensable
element of the topological order framework.

In the present paper we carry out such program from analyses of the dimerized $XY$
chain and the two-leg Heisenberg ladders with staggered and columnar dimerizations.
The first exactly-solvable model is equivalent to noninteracting Jordan-Wigner fermions.
To treat the ladders, we use the Jordan-Wigner tranformation and the mean-field approximation
to obtain an effective free-fermionic Hamiltonian for the spin ladder. In the earlier
related work it has been shown that such approximation is quite adequate, even quantitatively.
\cite{GYC11,GYC08,GYC07}  By utilizing duality transformations we demonstrate that it is
possible to map the problem of calculation of nonlocal SOPs in the chain and ladders onto the problem
of a local (Landau) order parameter in the dual representation, where the dual models
are given by exactly-solvable Hamiltonians and their order parameters can be calculated exactly.
It is then also straightforward to relate appearance of the SOP with the spontaneous symmetry
breaking in terms of the dual Hamiltonian.
In addition we show that the gapped phases not only have
distinct SOPs, but they can also be distinguished by the winding (Pontryagin) number.

These results allow us to stress close similarities between various gapped phases occurring in spin
models and topological insulators/supecronductors. The latter are of great current
interest. \cite{TI,FradkinBook13} The proposed approach provides a unifying theoretical framework
to deal with nonlocal order parameters in such seemingly different systems within the Landau paradigm.

The rest of the paper is organized as follows: In Sec.~\ref{DGXY} we
introduce and discuss the dimerized 1D $XY$ model. Using the duality transformation, the
SOPs and local order parameters are calculated exactly. We also find the winding numbers
for different phases of the model.  In Sec.~\ref{2LL} we present analogous results for the dimerized
two-leg ladders. A mean-field approximation is used to construct the effective Hamiltonian for the
Jordan-Wigner fermions which were mapped back onto exactly solvable (dual) spin models. The Appendix
contains details of the order parameter calculation for the quantum Ising chain with the three-spin
interaction which seems was never done before. The results are summarized and discussed in the
concluding Sec.~\ref{Concl}.

%
%
\section{Dimerized $XY$ Chain}\label{DGXY}
%
%
%
\subsection{Transformations of the Hamiltonian}
%
%
To analyze in depth various aspects of the conventional Landau and
the topological orders we start from the dimerized quantum $XY$ chain:
\begin{eqnarray}
\label{dgHam}
   H &=&\sum_{i=1}^{N}~  \frac{J}{4}  \big[ (1+\gamma)
 \sigma_{i}^{x}\sigma_{i+1}^{x}+ (1-\gamma) \sigma_{i}^{y}\sigma_{i+1}^{y}
 \nonumber \\
 &+& \delta
 (-1)^{i}(\sigma_{i}^{x}\sigma_{i+1}^{x}+\sigma_{i}^{y}\sigma_{i+1}^{y}) \big]+
 \frac12 (-1)^{i} h \sigma_{i}^{z} ~,
\end{eqnarray}
where $\sigma$-s are the standard Pauli matrices, $J$ is the nearest-neighbor
exchange coupling (we assume it to be antiferromagnetic), and $\gamma$ and
$\delta$ are the dimensionless parameters of anisotropy and dimerization,
respectively. This exactly-solvable model was first introduced and analyzed by
Perk \textit{et al}\cite{Perk75} in the presence of uniform and alternating
magnetic fields. (See also \cite{DelGamMod} for more recent work.) To
make connections with ladder models discussed below we take the Hamiltonian
(\ref{dgHam}) with an alternating transverse field. The standard Jordan-Wigner transformation (JWT)
\cite{Lieb61} maps spins onto free fermions
\begin{equation}
\label{Hspinor}
 H= \frac12 \sum_{k}\Psi^{\dag}_{k}\mathcal{H}(k) \Psi_{k}~,
\end{equation}
where the Fourier transforms of the Jordan-Wigner fermions residing on the
even/odd sites of the chain are unified in the Nambu spinor
\begin{equation}
  \Psi_{k}^{\dag}=\left(d_{e}^{\dag}(k),
  d_{o}^{\dag}(k),d_{e}(-k),d_{o}(-k)\right)~,
\label{spinor1}
\end{equation}
with the wavenumbers restricted to the Brillouin zone (BZ) $k \in
[-\pi/2,\pi/2]$ and we set the lattice spacing $a=1$. The $4\times4$ Hamiltonian
can be written as
\begin{equation}
\label{Hgamma}
  \mathcal{H}(k)= J  \cos k \Gamma_3 + J \delta \sin k \Gamma_4+
  h \Gamma_5 -J \gamma \sin k \Gamma_{13}~,
\end{equation}
where the five Dirac matrices are:
\begin{equation}
\label{Dirac}
\Gamma_{1,..,5}=\{\sigma_1 \otimes \mathbb{1},\sigma_2 \otimes \mathbb{1},
\sigma_{3} \otimes \sigma_{1},\sigma_{3} \otimes \sigma_{2},\sigma_{3} \otimes \sigma_{3} \}
\end{equation}
and $\Gamma_{13}=[\Gamma_1,\Gamma_3]/2i  =-\sigma_{2} \otimes \sigma_{1}$. This
Hamiltonian has four eigenvalues $\pm\epsilon^{\pm}$ where
\begin{equation}
\label{EpsMF}
 \epsilon^{\pm}(k)=J
 \sqrt{ \cos^2 k+\Big( \sqrt{\Big( \frac{h}{J} \Big)^2+ \delta^2 \sin^2 k}\pm
 \gamma \sin k \Big)^2}~.
\end{equation}
From (\ref{EpsMF}) we infer a remarkable property of the model
(\ref{dgHam}): each of the relevant perturbations $h, \delta, \gamma$ makes
the uniform isotropic $XX$ chain gapfull with the gap
\begin{equation}
 \Delta = J \big| \sqrt{\left(h/J\right)^2+\delta^2} \pm \gamma \big|~,
 \label{GapMF}
\end{equation}
while those terms cancel along the lines of quantum criticality
\begin{equation}
 \label{CritMF}
    \gamma=\pm\sqrt{\left(h/J\right)^2+\delta^2}~.
\end{equation}
and the model becomes gapless. The same cancelation effect happens in dimerized ladders
considered in the next section, but contrary to ladders, the present model can
be treated exactly at each step.

When we turn off the field and restrict ourselves to two relevant perturbations, pertinent
to other systems considered in this work, the spectrum of the model (\ref{dgHam})
becomes:\cite{DelGamMod}
\begin{equation}
\label{HdgSpectrum}
 \epsilon^{\pm}(k) = J \sqrt{\cos^{2}k+(\delta \pm \gamma)^{2}\sin^{2}k}
\end{equation}
with the gap
\begin{equation}
\label{GapXY}
  \Delta = J |\gamma \pm \delta|~.
\end{equation}
Now we will analyze the nature of order and symmetry changes on the lines of
the quantum phase transition $\gamma= \pm \delta$ on the parameter plane
($\delta, \gamma$). We utilize the duality transformation:\cite{Fradkin78}
\begin{eqnarray}
  \sigma_{n}^{x} &=& \tau_{n-1}^{x}\tau_{n}^{x}
  \label{sigmaX} \\
  \sigma_{n}^{y} &=& \prod_{l=n}^{N} \tau_{l}^{z}~,
  \label{sigmaY}
\end{eqnarray}
where $\tau$-s obey the standard algebra of the Pauli operators, and they
reside on the sites of the dual lattice, which can be placed between the sites
of the original chain where the operators $\sigma$ reside. In terms of the dual
operators, the Hamiltonian (\ref{dgHam}) becomes a sum of two completely
decoupled 1D Quantum Ising Model (QIM) Hamiltonians defined on the even and odd
sites of the dual lattice:
\begin{eqnarray}
  H &=& H_{\mathrm{even}} + H_{\mathrm{odd}}  \nonumber \\
  H_{\mathrm{even}}&=&  \frac{1}{4}\sum_{l=1}^{N/2}
   \big(  J^{+-} \tau_{2l-2}^{x} \tau_{2l}^{x}+J^{-+}\tau_{2l}^{z}\big)   \label{He} \\
  H_{\mathrm{odd}}&=&  \frac{1}{4}\sum_{l=1}^{N/2}
   \big(  J^{++} \tau_{2l-1}^{x} \tau_{2l+1}^{x}+J^{--}\tau_{2l-1}^{z}\big)~,  \label{Ho}
\end{eqnarray}
with the notations
\begin{equation}
\label{Jpm}
 J^{\pm\pm} =  J(1\pm\gamma\pm\delta).
\end{equation}
%
%
\subsection{Local and String Order Parameters}
%
%
We define the string operator as
\begin{equation}
\label{String}
 O^{\alpha}_m=
 \exp \Big[\frac{i\pi}{2}\sum_{k \leftarrowtail m}\sigma_{k}^{\alpha} \Big]~,
\end{equation}
where $\alpha = x, y, z$ and the summation is carried out along all sites of the string, left from the $m$-th site.
In case of the chain model this is unambiguous and means $k<m$, while for ladders or other models the path of
the string must be specified, as we discuss in the following sections. The string order
parameter (SOP) $\mathcal{O}_{\alpha}$ is determined from the limit $n-m \to \infty$ of the string-string
correlation function
\begin{equation}
\label{S-S}
 \langle O^{\alpha}_m   O^{\alpha}_n \rangle =
   (-1)^{m-1} \Big< \exp \Big[\frac{i\pi}{2}\sum_{k=m}^{n-1}\sigma_{k}^{\alpha} \Big] \Big>~.
\end{equation}
Taking $m=1$ and $n=2l+1 $ in  \eqref{S-S} (note that SOP for odd number of spins vanishes due to symmetry), the SOP
is introduced as
\begin{equation}
\label{SOPDef}
 \mathcal{O}^2_{\alpha} =\lim_{l\rightarrow\infty}(-1)^{l}
 \Big\langle \prod_{k=1}^{2l} \sigma_{k}^{\alpha} \Big\rangle~.
\end{equation}
In the original proposal by den Nijs and Rommelse\cite{denNijs89} and in the
subsequent work on the spin chains (see, e.g., \cite{HiddenSSB,Hida92}) the SOP
was identified (up to some minor variations) with the limit of the string-string  correlation function,
which is not convenient, since such SOP has a wrong dimension of square of the order parameter.
The definition we use here \cite{Berg08} is more consistent with the standard theory of critical phenomena
and is in line with the definition of the Landau order parameter via a correlation function of local operators.
We will show that in the critical region  $ \mathcal{O}_{\alpha} \propto |t|^{\beta}$, where $t$ is a distance
from the critical point and the critical index of the order parameter $\beta$
correctly enters all the scaling relations.

Using the duality transformation (\ref{sigmaX}) in
Eq.~(\ref{SOPDef}) we get
\begin{equation}
\mathcal{O}_x^2 =
\lim_{l\rightarrow\infty}(-1)^{l}\left<\tau_{0}^{x}\tau_{2l}^{x}\right>~.
\label{SOPdual}
\end{equation}
So, the nonlocal SOP defined on the sites of the direct lattice becomes a local
order parameter on the dual lattice. This implies that for the
phase transitions with nonlocal orders the conventional Landau framework
can be recovered via duality.

\begin{figure}[h]
\centering{\includegraphics[width=8 cm] {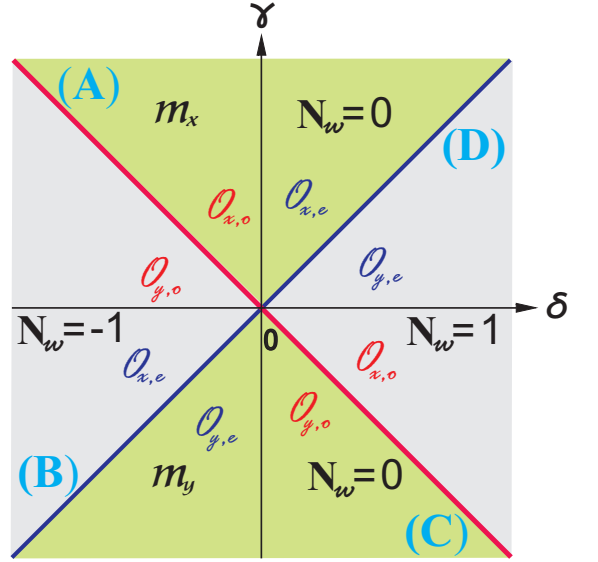}}
 \caption{(Color online) Phase diagram of the anisotropic dimerized XY chain.
Nonvanishing string and local order parameters, topological winding numbers
are shown in four sectors A,B,C,D of the ($\delta, \gamma$) parametric plane.
The violet/red lines $\gamma=\pm\delta$ are the lines of quantum phase transitions
(gaplessness).} \label{PhaseDiagXY}
\end{figure}

According to the classical results \cite{Pfeuty70,Barouch71} for the 1D QIM
with the Hamiltonian
\begin{equation}
 \label{HQIM}
H =  \sum_{i=1}^{N}\left[ J \tau_{i}^{x} \tau_{i+1}^{x} + h
\tau_{i}^{z}\right]~,
\end{equation}
the model is disordered at $\lambda \equiv J/h \leq 1$, while the long-range
order  $m_{x} \neq 0$ appears at $\lambda >1$. The order parameter is
obtained from the correlation function:
\begin{equation}
\lim_{l\rightarrow\infty}\left<\tau_{i}^{x}\tau_{i+l}^{x}\right> =
(-1)^{l}\left(1-\lambda^{-2}\right)^{\frac{1}{4}}=(-1)^{l}m_{x}^{2}~.
\label{QIMCorr}
\end{equation}
From comparison of the Hamiltonians (\ref{He},\ref{Ho},\ref{HQIM}) we see that
if $\lambda_\sharp>1$, where
\begin{equation}
\lambda_{e/o} =\frac{1 + \gamma \mp \delta}{1-  \gamma \pm \delta}~,
\label{lambda}
\end{equation}
then $\tau^x$ is ordered on either even or odd dual sublattices. So we should
distinguish between the even and odd SOP defined on the sites of the even or odd dual
sublattices, respectively. As one can see, the SOP defined by Eq.~(\ref{SOPdual})
is even, and we will denote it by $\mathcal{O}_{x,e}$ from now on.
To define the odd SOP ($\mathcal{O}_{x,o}$) we can take $m=2$ and $n=2l+2$ in Eq.~(\ref{S-S}),
then $\mathcal{O}_{x,o}$ is given by Eq.~(\ref{SOPdual}) where the correlation function
on its r.h.s is now $ \langle \tau_{1}^{x}\tau_{2l+1}^{x} \rangle$. Using the above formulas
the even and odd SOP can be calculated exactly:
\begin{equation}
\mathcal{O}_{x,e/o} =
\left\{
\begin{array}{lr}
2^{1/4} \big( \frac{t_{\mp}}{(1+t_{\mp})^2} \big)^{1/8}
& t_{\mp} \geq 0 \\
0& t_{\mp} < 0
\end{array}
\right. \label{Oxeo}
\end{equation}
where we denote
\begin{equation}
\label{tpm}
  t_{\pm} = \gamma \pm \delta
\end{equation}
Let us now calculate the $y$ component of the SOP. The duality transformations
(\ref{sigmaX},\ref{sigmaY}) with the interchange $x \leftrightarrow y$ map
the Hamiltonian (\ref{dgHam}) again onto a sum of two even and odd QIM Hamiltonians:
\begin{eqnarray}
  H_{\mathrm{even}}&=&  \frac{1}{4}\sum_{l=1}^{N/2}
   \big(  J^{-+} \tau_{2l-2}^{y} \tau_{2l}^{y}+J^{+-}\tau_{2l}^{z}\big)   \label{HeY} \\
  H_{\mathrm{odd}}&=&  \frac{1}{4}\sum_{l=1}^{N/2}
   \big(  J^{--} \tau_{2l-1}^{y} \tau_{2l+1}^{y}+J^{++}\tau_{2l-1}^{z}\big)  \label{HoY}
\end{eqnarray}
Following the same steps as above we easily find:
\begin{equation}
\mathcal{O}_{y,e/o} =
\left\{
\begin{array}{lr}
0 & t_{\mp} \geq 0 \\
\mathcal{O}_{x,e/o}(-t_{\mp}) & t_{\mp} < 0
\end{array}
\right. \label{Oyeo}
\end{equation}
We show the phase diagram of the dimerized $XY$ model in
Fig.~\ref{PhaseDiagXY}. The two lines of quantum phase transitions $\gamma =
\pm \delta$ divide the parameter plane into four regions denoted by A-D. For
each region we indicate nonvanishing order parameters which characterize
different phases located there.

The next important step in our analysis is to establish relation between the
conventional long-range order with local order parameter(s) and the nonlocal
(topological) SOPs. Using the duality transformation (\ref{sigmaX}) and
decoupling of the dual Hamiltonian into the even and odd terms, we find the
magnetization $m_x$ of the spins of the original Hamiltonian (\ref{dgHam}) as:
\begin{equation}
 \label{Mx}
  m_{x}^{2} = \lim_{l\rightarrow\infty}\langle\sigma_{1}^{x}\sigma_{2l}^{x}\rangle
  = \lim_{l\rightarrow\infty}\langle(\tau_{0}^{x}\tau_{2l}^{x})
  (\tau_{1}^{x}\tau_{2l-1}^{x})\rangle~.
\end{equation}
The $y$ component of the magnetization can be treated in a similar way, so we
get
\begin{equation}
\label{MO}
    m_\alpha= \mathcal{O}_{\alpha,o} \mathcal{O}_{\alpha,e}~,~~\alpha=x,y.
\end{equation}
As we see from the model's phase diagram in Fig.~\ref{PhaseDiagXY}, the even
and odd SOPs are mutually exclusive in the regions B and D. Those are the
regions without conventional local (Landau) order. The phases in the regions B
and D possess only a nonlocal (topological) order which is quantified by the
SOP. As we conclude from the duality mapping, the dimerized $XY$ model
(\ref{dgHam}) possesses the hidden $\mathbb{Z}_2 \otimes \mathbb{Z}_2$ symmetry
which was also noticed earlier in various spin chains and
ladders.\cite{HiddenSSB} In the present analytically solvable model the origin
of such symmetry can be easily traced to the $\mathbb{Z}_2$ symmetry of the
even and odd Ising models (\ref{He},\ref{Ho}). Nonzero even SOP
$\mathcal{O}_{x,e}$ and odd SOP $\mathcal{O}_{x,o}$ in the sectors B and D are
due to spontaneous breaking of the $\mathbb{Z}_2$ symmetry in the even
(\ref{He}) or odd (\ref{Ho}) sectors of the dual Hamiltonian, respectively.
These SOPs signal appearance of the spontaneous ``dual" magnetization $\langle
\tau^x_e \rangle$ or $\langle \tau^x_o \rangle$. We have to emphasize an
important issue: since mappings of the original model (\ref{dgHam}) onto the
dual models (\ref{He},\ref{Ho}) or (\ref{HeY},\ref{HoY}) are results of the
\textit{different} dual transformations, the analyses in terms of the SOP
$\mathcal{O}_x$ or $\mathcal{O}_y$ are \textit{complimentary}, and the order
parameters $\mathcal{O}_x$ and  $\mathcal{O}_y$, even if they both are found to
be nonzero in some parts of the phase diagram should not be understood as
\textit{coexistent}.

We conclude from Eqs.~(\ref{Oxeo},\ref{Oyeo}) (see also Fig.~\ref{PhaseDiagXY})
that contrary to common belief, the even and odd SOPs can coexist. In the
region A of the phase diagram which corresponds to $\gamma^2 > \delta^2$,
$\gamma>0$ both even and odd SOPs $\mathcal{O}_{x,e/o}$ are nonvanishing, and
then, as follows from Eq.~(\ref{MO}) the local order (magnetization) $m_x \neq
0$ appears in the original model (\ref{dgHam}). Similarly, when $\gamma^2 >
\delta^2$, $\gamma<0$ (region C of the phase diagram), the magnetization $m_y
\neq 0$. The appearance of the local order, or, equivalently for this model,
the coexistence of the even and odd SOPs are the result of the spontaneous
$\mathbb{Z}_2 \otimes \mathbb{Z}_2$ symmetry breaking simultaneously in both
even and odd sectors of the dual Hamiltonian. From
Eqs.~(\ref{Oxeo},\ref{Oyeo},\ref{MO}) we find the numerical values of the local
order parameters:
\begin{equation}
 \label{Mxy}
    m_x(\gamma)=m_y(-\gamma)=
    \sqrt{2 }\Big( \frac{\gamma^2-\delta^2}{((1+\gamma)^2-\delta^2 )^2}
    \Big)^{1/8}~.
\end{equation}
In the limit $\delta =0$ the above equation coincides with the result of
Barouch and McCoy.\cite{Barouch71}

Another useful limit of the model (\ref{dgHam}) is $\gamma=0$ when it becomes a
dimerized $XX$ chain. This model does not have a local long-range order, and
its phase diagram can be read from the line $\gamma =0$ in
Fig.~\ref{PhaseDiagXY}. The topological order in the phases $\delta >0$ or
$\delta <0$ is characterized by a pair of mutually exclusive SOPs
$\mathcal{O}_{\alpha ,o}$ and $\mathcal{O}_{\alpha,e}$  (we can choose
$\alpha=x$ or $\alpha=y$)  separated by a quantum critical point $\delta=0$
where all SOPs vanish. From (\ref{Oxeo},\ref{Oyeo}) we find
\begin{equation}
   \label{SOPXX}
   \mathcal{O}_{x,o}= \mathcal{O}_{y,e}=
   2^{1/4} \Big( \frac{\delta}{(1+\delta)^2} \Big)^{1/8}~,~~ \delta
   >0~.
\end{equation}
The magnitudes of the SOPs are symmetric with respect to $\delta$ and
$\mathcal{O}_{x,e}(\delta) = \mathcal{O}_{y,o}(\delta)=
\mathcal{O}_{x,o}(-\delta)$ for $\delta <0$.

If we define a primary order parameter as the one which is nonzero exclusively
for a given phase, then we can identify $m_x$ and $m_y$ as the primary order
parameters for the magnetized phases residing, respectively, in the sectors A
and C of the phase diagram. The topological order of the phases B and D is
characterized by the pair of the primary (string) parameters
$\mathcal{O}_{x,e}$ and $\mathcal{O}_{x,o}$ for the choice
(\ref{sigmaX},\ref{sigmaY}) of the dual transformation. (If we swap $x
\leftrightarrow y$, we end up with the dual representation where the primary
order parameters are $\mathcal{O}_{y,e}$ and $\mathcal{O}_{y,o}$.) From
behavior of the gap (\ref{GapXY}) and the primary order parameter (cf.
(\ref{Oxeo},\ref{Oyeo},\ref{Mxy}), whichever applicable) near the lines $\gamma
= \pm \delta$ of the quantum phase transition we find the critical indices
of the order parameter $\beta=1/8$ (cf. its definition below Eq.~(\ref{SOPDef}))
and of the correlation length $\nu=1$. (The latter can be read from the gap
equation (\ref{GapXY}) since $\xi^{-1} \propto \Delta \propto |t|^\nu$).
So the critical behavior of the model on the phase
boundaries belongs to the 2D Ising universality class as it must, since the
Hamiltonian (\ref{dgHam}) is equivalent to free fermions.
%
%
\subsection{Topological Winding Numbers}
%
%
In recent years it has been proven that many quantum phase transitions with
hidden orders are accompanied by a change of topological numbers
\cite{FradkinBook13,VolovikBook,TI,Hatsugai-1,Hatsugai-2,Ezawa13,Mila13}. In this section we
calculate the winding number (or the Pontryagin index) in the dimerized $XY$
spin chain to characterize its different gapped phases. These topological
numbers were calculated recently in similar 1D
systems.\cite{Wu12,Niu12,Ezawa13} Following the formalism of
Ref.[\onlinecite{SchnyderRyu11}] we first rewrite (\ref{Hgamma}) as the
Bogoliubov-de Gennes Hamiltonian of the topological superconductor (class
DIII):
\begin{equation}
\label{BdG}
  \mathcal{H}(k)=\left(%
\begin{array}{cc}
  \hat{\mathfrak{h}}(k) & \hat{\Delta}(k) \\
  \hat{\Delta}^\dag (k) & -\hat{\mathfrak{h}}(k) \\
\end{array}%
\right)~,
\end{equation}
where
\begin{eqnarray}
   \hat{\mathfrak{h}}(k) &\equiv&  J \cos k \sigma_1+   J \delta \sin k \sigma_2
   + h \sigma_3~, \label{hmat} \\
   \hat{\Delta}(k)   &\equiv &    -i J \gamma \sin k \sigma_1 ~. \label{Dmat}
\end{eqnarray}
By a unitary transformation the above Hamiltonian can be brought to the block
off-diagonal form
\begin{equation}
\label{HD}
  \tilde{\mathcal{H}}(k)=\left(%
\begin{array}{cc}
  0 & \hat{D}(k) \\
  \hat{D}^\dag (k) & 0 \\
\end{array}%
\right)~,
\end{equation}
with the operator
\begin{equation}
\label{Dk}
    \hat{D}(k)=\hat{\mathfrak{h}}(k)+\hat{\Delta}(k)~,
\end{equation}
which has two eigenvalues $\pm \lambda(k)$ with
\begin{equation}
\label{lambdak}
    \lambda(k)=J \sqrt{(h/J)^2+1+ (\delta^2-\gamma^2-1) \sin^2 k-i \gamma \sin 2k
    }~.
\end{equation}
Note a useful relation between the eigenvalues of $\hat{D}(k)$,
$\hat{D}^\dag(k)$ and of the Hamiltonian (\ref{EpsMF}):
\begin{equation}
 \label{EpsLam}
    \epsilon^+ (k) \epsilon^- (k) = \lambda (k) \lambda^* (k)
\end{equation}
In one spatial dimension the winding number defined as \cite{SchnyderRyu11}
\begin{equation}
\label{Nw}
  N_w^r=\frac{1}{4\pi i} \int_{BZ} dk \mathrm{Tr}[\partial_k \ln \hat{D}-
         \partial_k \ln \hat{D}^\dag ]
\end{equation}
can be readily calculated analytically for this model. We set $h=0$ and
calculated the winding number in all four regions of the phase diagram in
Fig.~\ref{PhaseDiagXY}. To make a connection between the definition (\ref{Nw})
and a more intuitive definition of the winding number let us introduce a
two-component unit vector $n(k)=\left(n_1(k),n_2(k)\right)$ constructed from
the spectrum of the model. The eigenvalues (\ref{HdgSpectrum}) allow us to
define two unit vectors with the components
\begin{eqnarray}
  n_1(\epsilon^\pm(k)) &=& J \cos k/ \epsilon^\pm(k)~, \label{n1}\\
  n_2(\epsilon^\pm(k)) &=&  J (\delta \pm \gamma) \sin k/ \epsilon^\pm(k)~,\label{n2}
\end{eqnarray}
which can also be related to the Bogoliubov angle $e^{i \theta_k}=n_1(k)+i n_2(k)$. (For the
definition of  $\theta_k$ see, e.g., Ref.~\cite{Abanov05}.)
The winding number defined as\cite{FradkinBook13,VolovikBook}
\begin{equation}
\label{Npm}
 N_{\pm}=\frac{1}{2\pi} \sum_{i,j=1}^2 \int_{BZ} dk \varepsilon_{ij} n_i
 (\epsilon^\pm (k))\partial_k  n_j(\epsilon^\pm (k))
\end{equation}
counts the number of loops wrapped by the unit vector around the origin while
the wavevector spans over the Brillouin zone. One can show that the definition
(\ref{Nw}) yields
\begin{equation}
\label{Nwpm}
  N_w^r=N_- - N_+= \frac12 \big[ \mathrm{sign}(\delta-\gamma)-\mathrm{sign}(\delta+\gamma) \big]~.
\end{equation}
The topological number $N_w^r$ accounts for the ``relative" winding of two
normalized eigenvalues of the Hamiltonian. We find the winding number $N_w$
which adds up the loops made by each of the eigenvectors more convenient for
various applications:
\begin{equation}
\label{Nwsum}
  N_w=N_- + N_+= \frac12 \big[ \mathrm{sign}(\delta-\gamma)+\mathrm{sign}(\delta+\gamma) \big]~.
\end{equation}
As one check from Eqs. (\ref{Nwpm}) or (\ref{Nwsum}), both topological numbers
change by $\pm 1$ on the phase boundaries $\gamma = \pm \delta$, signalling the phase transition.
We choose $N_w$ as a complimentary topological order parameter characterizing a given phase, since its
behavior is in accord with the standard wisdom: as one can see in Fig.~\ref{PhaseDiagXY},
$N_w=0$ in the topologically trivial regions where a conventional local order exists.

%
%
\section{Dimerized Two-leg Ladder}\label{2LL}
%
%
%
\subsection{Model and Effective Mean-Field Hamiltonian}
\begin{figure}[h]
\epsfig{file=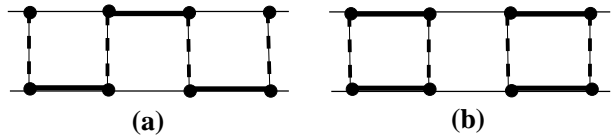,width=0.45\textwidth,angle=0} \caption{Dimerized
two-leg ladder. Bold/thin/dashed lines represent the stronger/weaker
chain coupling $J(1 \pm \delta)$ and rung coupling $J_\bot$,
respectively.  Dimerization patterns: (a) - staggered; (b)-
columnar.} \label{Dim_Lad}
\end{figure}

%
%
Now we present the results for a two-legged ladder with intrinsic
dimerization. In line with the earlier work\cite{GYC11,GYC08,GYC07}
we consider the two possible dimerization patterns of the ladder: staggered
and columnar, shown in Fig.~\ref{Dim_Lad}.
%
%
The Hamiltonian of the dimerized spin-$\frac12$ ladder with two legs is given by:
\begin{eqnarray}
\label{Ham}
 H &=& \sum_{n=1}^{N}\sum_{\alpha=1}^{2} J_{\alpha}(n)
 \mathbf{S}_{\alpha}(n) \cdot \mathbf{S}_{\alpha}(n+1) \nonumber \\
 &+& J_\bot \sum_{n=1}^{N} \mathbf{S}_{\alpha}(n) \cdot \mathbf{S}_{\alpha+1}(n),
\end{eqnarray}
where the dimerization occurs along the chains ($\alpha=1,2$) only,
with the rung coupling $J_\bot$ taken as constant. All the spin exchange couplings
are antiferromagnetic. The dimerization patterns are then defined as:
\begin{eqnarray}
J_{\alpha}(n)=J[1-(-1)^{n+\alpha}\delta] & \mbox{(staggered)} \label{staggered}\\
J_{\alpha}(n)=J[1-(-1)^{n}\delta] & \mbox{(columnar)} \label{columnar}
\end{eqnarray}
with periodic boundary conditions along the chains and open boundary conditions
in the rung directions.

Critical properties of dimerized ladders are known from analytical and
numerical studies:\cite{Delgado,Kotov99,Cabra99,Nersesyan00,Okamoto03,Nakamura03,GYC11,GYC08,MoreDimLadd}
the staggered ladder can undergo a continuous quantum phase transition and is gapless
on the line of quantum criticality, while the columnar
ladder is always gapped and does not undergo any transition. We infer also from our earlier
work\cite{GYC11,GYC08} that the mean-field approximation for this model is quite adequate, even
quantitatively. So we apply this approximation to map the spin ladder model onto a problem
of effective quadratic fermionic Hamiltonian. The latter will be utilized to study analytically
the critical properties of two dimerizations, calculation of SOPs and topological numbers, similarly
to what was done above for the exactly solvable chain.

\begin{figure}[h]
\includegraphics[width=6.0cm]{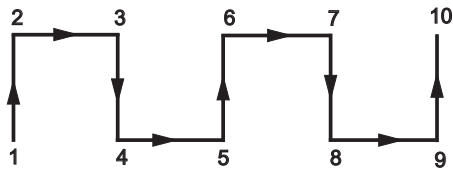}
\caption{Path of the Jordan-Wigner transformation used for the two-leg ladder.}
\label{JWpath}
\end{figure}

We map the spin ladder Hamiltonian (\ref{Ham}) onto fermions via a JWT. There are different ways to introduce 
this transformation when we depart from a single-chain problem (see, e.g., 
Refs.[\onlinecite{Azzouz,Dai98Etal,Xiang07}]). We chose the snake-like path for the JWT used in \cite{Dai98Etal}. 
Labelling the sites of the ladder along the path as shown in
Fig.~\ref{JWpath}, the JWT is defined as
\begin{equation}
\label{JWT}
\sigma_{n}^{+}=c_{n}^\dag  \exp \Big( i\pi\sum_{l=1}^{n-1}
c^\dag_l c_l \Big)
\end{equation}
and $\sigma_n^z= 2c^\dag_n c_n-1$,  where $\sigma_{n}^\pm =\frac12 (\sigma_n^x \pm i \sigma_n^y)$.

%
%
We treat the interaction terms via a standard mean-field decoupling:
\begin{equation}
\label{MFA}
  \Big(\hat{n}_k -\frac12 \Big) \Big(\hat{n}_m -\frac12 \Big) \approx \frac14
  -\big( \langle c_m^\dag c_k \rangle c_k^\dag c_m +h.c. \big) + |\langle c_m^\dag c_k \rangle |^2
\end{equation}
Then the spin ladder (\ref{Ham}) maps onto the following fermionic Hamiltonian:
\begin{widetext}
\begin{equation}
\label{Hmf1}
  H_{\mathrm{MF}}= N\mathcal{C}+\frac12 \sum_{m,\alpha} J_\alpha(m)
  \big[ e^{i \hat{\Phi}_\alpha(m)}+2 t_{\alpha \parallel}(m) \big] c_\alpha^\dag(m)c_\alpha(m+1)+
  \frac12 J_\bot \sum_m \big[ 1+2 t_\bot(m) \big] c_1^\dag(m)c_2(m)+h.c.
\end{equation}
\end{widetext}
Although we use the JWT path as shown in Fig.~\ref{JWpath}, we keep the labelling of fermions residing
on the ladder in agreement with the notations of the original spin Hamiltonian (\ref{Ham}).
The phase entering Eq.~(\ref{Hmf1}) is defined as
\begin{equation}
 \hat{\Phi}_\alpha(m)=
\left\{
\begin{array}{lr}
0~ &\alpha=1,~ m=2l \\
\pi (\hat{n}_2(m)+\hat{n}_2(m+1) )~ &\alpha=1,~ m=2l-1 \\
0~ &\alpha=2,~ m=2l-1 \\
\pi (\hat{n}_1(m)+\hat{n}_1(m+1) )~ &\alpha=2,~ m=2l
\end{array}
\right.
\label{Phase}
\end{equation}
and the effective hopping terms are
\begin{eqnarray}
 t_{\alpha \parallel}(m) &=&
 \langle c_\alpha(m) c_\alpha^\dag (m+1) \rangle    \label{tpar}\\
  t_\bot(m) &=&
 \langle c_1(m) c_2^\dag (m+1) \rangle \label{tperp}
\end{eqnarray}
Note that the coupling $J_\bot$ spoils the commutation of $\hat{\Phi}_\alpha(m)$ and
$H_{\mathrm{MF}}$. We however will approximately treat the phase as a good quantum number,
i.e. $\hat{\Phi}_\alpha(m) \approx \Phi_\alpha(m)$. The quantitative validity of this approximation 
was confirmed in earier related work \cite{Azzouz,GYC07,GYC08,Toplal14}. Then according to the Lieb
theorem\cite{Lieb94} the ground state of this quadratic Hamiltonian is in the $\pi$-flux
phase,\cite{Affleck88} which amounts to
\begin{equation}
\label{Piflux}
  e^{i \Phi_\alpha(m)}+2 t_{\alpha \parallel}(m)=(-1)^{m+\alpha-1} (1+2 t_{\alpha \parallel})~.
\end{equation}
Assuming further $t_{\alpha \parallel} \approx t_\parallel$ and $t_\bot(m) \approx t_\bot$ we get
\begin{widetext}
\begin{equation}
\label{Hmf2}
  H_{\mathrm{MF}}= N\mathcal{C}+\frac12 \sum_{n,\alpha} (-1)^{n+\alpha-1} J_{\alpha R}(n)
   c_\alpha^\dag(n)c_\alpha(n+1)+
  \frac12 J_{\bot R} \sum_n  c_1^\dag(n)c_2(n)+h.c.
\end{equation}
\end{widetext}
where $J_{\alpha R}(n)$ is given by Eqs.(\ref{staggered},\ref{columnar}) with $J \mapsto J_R$, and the renormalized couplings are
\begin{eqnarray}
J_R &=& J (1+2 t_\parallel) \label{JR}\\
J_{\bot R} &=& J_\bot (1+2 t_\bot)~. \label{JT}
\end{eqnarray}
The constant term
\begin{equation}
\label{C}
  \mathcal{C}= \frac14 (J_\bot + 2J) +2 J t_\parallel^2 +J_\bot t_\bot^2~.
\end{equation}
The renormalization parameters $t_\parallel$ and $t_\bot$ must be determined self-consistently from minimization of the mean-field Hamiltonian (\ref{Hmf2}). In fact this was done before in a slightly more sophisticated mean-field treatment \cite{GYC08} with $t_{1 \parallel} \neq t_{2 \parallel}$. (For earlier work on this mean-field approach, see \cite{Azzouz}.) The present mean-field Hamiltonian gives essentially the same results for the ground state energies, eigenvalues, gaps, renormalization parameters, and the phase diagram for the both staggered and columnar ladders\cite{Toplal14} with some minor numerical differences. Keeping in mind that the mean-field parameters lie within the range $0 \leq (t_\parallel, t_\bot ) \leq \frac12$ for all possible values of the Hamiltonian's parameters, \cite{Toplal14} it suffices for the goals of the present study to express results directly in terms of $J_R$ and $J_{\bot R}$. So in the following these renormalized couplings are treated as free model parameters of the effective Hamiltonian \eqref{Hmf2}.
%
%
\subsection{Spectra and Winding Numbers}
%
%
%
%
\subsubsection{Staggered Ladder}
%
%
To further simplify the effective Hamiltonian (\ref{Hmf2}) one can perform a canonical transformation mapping $(-1)^n c_\alpha^\dag(n)c_\alpha(n+1) \mapsto c_\alpha^\dag(n)c_\alpha(n+1)$.
Introducing instead of $c_\alpha$ two distinct fermions residing on the even/odd sites $d_{\alpha, e/o}$ and Fourier transforming, we rewrite the Hamiltonian as
\begin{equation}
\label{HspLad}
 H_{\mathrm{MF}}=  N\mathcal{C}+  \sum_{k}\Psi^{\dag}_{k}\mathcal{H} (k) \Psi_{k}~,
\end{equation}
where the spinor
\begin{equation}
  \Psi_{k}^{\dag}=\left(d_{1,e}^\dag (k),d_{1,o}^\dag (k),d_{2,e}^\dag (k),d_{2,o}^\dag (k)\right)~,
\label{spinor2}
\end{equation}
and the wavenumbers restricted to the Brillouin zone $k \in [-\pi/2,\pi/2]$. The $4\times4$ Hamiltonian
of the staggered phase can be written as
\begin{equation}
\label{Hst}
  \mathcal{H}^{\mathrm{st}}(k)= \frac12 J_{\bot R} \Gamma_1+ J_R \cos k \Gamma_3+ J_R \delta \sin k \Gamma_{35}~,
\end{equation}
where the Dirac matrices are defined in (\ref{Dirac}) and $\Gamma_{35}=- \mathbb{1} \otimes \sigma_{2}$. This Hamiltonian has four eigenvalues $\pm\epsilon^{\pm}$ where
\begin{equation}
\label{EpsSt}
 \epsilon^{\pm}(k)=J_R
 \sqrt{ \cos^2 k+ \Big(\delta \sin k \pm \frac{J_{\bot R}}{2 J_R} \Big)^2}~.
\end{equation}
The spectrum has the gap
\begin{equation}
\label{Gap2L}
  \Delta = J_R \Big|\delta \pm \frac{J_{\bot R}}{2 J_R} \Big|~.
\end{equation}
Similar to the dimerized $XY$ chain, two relevant perturbations cancel on the lines of quantum critical transition
\begin{equation}
\label{CrLines}
  \frac{J_{\bot R}}{2 J_R} = \pm \delta ~,
\end{equation}
where the staggered two-leg ladder becomes gapless. (A more refined mean-field treatment of this model's phase diagram
and comparison to numerical results can be found in \cite{GYC08}. The present version of the mean field results in the 
numerically very accurate phase diagram \cite{Okamoto03}.)  
From comparison of the eigenvalues of the ladder's effective Hamiltonian (\ref{EpsSt}) and those of the $XY$ chain in the alternating transverse field (\ref{EpsMF}), we can infer the equivalence between the spectra of those models, and even match corresponding parameters:
\begin{eqnarray}
\underline{\mathrm{Staggered~ladder}}~&\Longleftrightarrow&~of the 
\underline{XY ~\mathrm{chain}~\eqref{dgHam}, ~\delta=0} \nonumber \\
J_R   ~&\longleftrightarrow&~ J \nonumber \\
\frac12 J_{\bot R} ~&\longleftrightarrow&~ h \nonumber \\
\delta ~&\longleftrightarrow&~ \gamma \nonumber
\end{eqnarray}
Recall also that the fermionized Hamiltonian of the $XY$ chain is equivalent to a topological superconductor (\ref{BdG}).
One can also establish equivalence and match parameters of the effective ladder Hamiltonian to those of the two-leg Kitaev ladder \cite{Xiang07} or to the $XY$ chain in a uniform transverse field and its dual model (for details of the latter, see next subsection). This is hardly surprising since all these models map onto quadratic fermionic Hamiltonians.

\begin{figure}[h]
\centering{\includegraphics[width=8 cm] {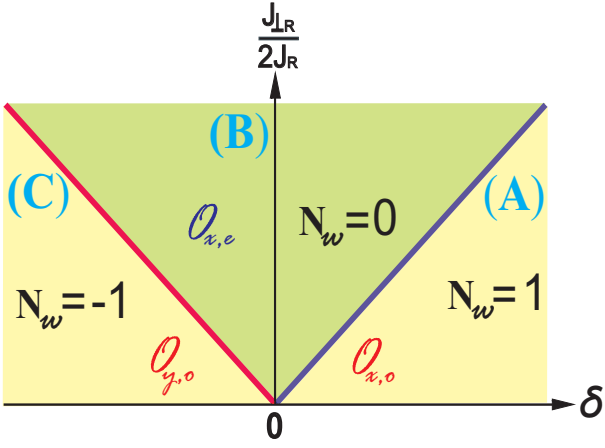}}
\caption{(Color online) Phase diagram of the two-leg staggered ladder in the parametric plane
($J_{\bot R}/2J_{R}, \delta$). The violet/red lines $J_{\bot R}/2J_{R} = \pm \delta$
are the lines of quantum phase transitions (gaplessness) separating two phases (B) and (A,C)
characterized by their distinct SOPs and winding  numbers.}
\label{PhaseDiagStag}
\end{figure}

To calculate the topological winding number $N_w$ in different phases of the staggered ladder we use
a unitary transformation to bring the Hamiltonian (\ref{Hst}) to the block diagonal form
\begin{equation}
\label{HstDiag}
  \tilde{\mathcal{H}}^{\mathrm{st}}(k)=\left(%
\begin{array}{cc}
  \hat{\mathfrak{h}}_+(k) & 0 \\
  0 &  \hat{\mathfrak{h}}_-(k) \\
\end{array}%
\right)~,
\end{equation}
familiar from the context of topological insulators and superconductors,\cite{TI} where  $\hat{\mathfrak{h}}_-(k)= \hat{\mathfrak{h}}_+^*(-k)$. Explicitly the operators
$\hat{\mathfrak{h}}_\pm (k)$ are
\begin{equation}
 \label{hpm}
 \hat{\mathfrak{h}}_\pm (k) = J_R  \cos k \sigma_1-
   \big( J_R \delta \sin k \pm \frac12 J_{\bot R} \big) \sigma_2~,
\end{equation}
and their eigenvalues are given by Eq.(\ref{EpsSt}). Using the eigenvalues (\ref{EpsSt}) to yield the unit vector (\ref{n1},\ref{n2}), we get from definition (\ref{Npm})
\begin{equation}
\label{NpmLad}
  N_\pm= \frac{1}{2 \pi} \int_{-\pi/2}^{\pi/2} dk \frac{\delta \pm \kappa \sin k}{\cos^2 k+(\delta \sin k \pm \kappa)^2}~,
\end{equation}
where we denote
\begin{equation}
\label{kappa}
  \kappa \equiv \frac{J_{\bot R}}{2 J_R}~.
\end{equation}
One can check that $N_+=N_-$. The result for the winding number $N_w$ can be cast in a simple form:
\begin{equation}
\label{NwsumLad}
  N_w=N_- + N_+= \frac12 \big[ \mathrm{sign}(\delta-\kappa)+\mathrm{sign}(\delta+\kappa) \big]~.
\end{equation}
We have shown the winding numbers for each phase on the phase diagram of the staggered ladder in Fig.~\ref{PhaseDiagStag}.  (Recall that
we analyze the ladder with antiferromagnetic couplings.) The complimentary order parameter $N_w$ changes by $\Delta N_w= \pm 1$ on the phase boundaries $J_{\bot R}/2 J_R = \pm \delta$. The sectors (A) and (C) correspond to the same topologically nontrivial phase.

%

%
%
%
\subsubsection{Columnar Ladder}
%
%
The $4\times4$ Hamiltonian of the columnar phase is
\begin{equation}
\label{Hcol}
  \mathcal{H}^{\mathrm{col}}(k)= \frac12 J_{\bot R} \Gamma_1+ J_R \cos k \Gamma_3+ J_R \delta \sin k \Gamma_{4}~,
\end{equation}
and it has two two-fold degenerate eigenvalues $\pm\epsilon$ where
\begin{equation}
\label{EpsCol}
 \epsilon(k)=J_R
 \sqrt{ \cos^2 k+ \delta^2 \sin^2 k + \Big( \frac{J_{\bot R}}{2 J_R} \Big)^2}~.
\end{equation}
Contrary to the staggered case, the columnar ladder is always gapped with the gap
\begin{equation}
\label{Gap2LCol}
  \Delta = J_R \sqrt{ \delta^2 + \Big( \frac{J_{\bot R}}{2 J_R} \Big)^2}~,
\end{equation}
and no phase transition occurs on the whole plane $(\delta, J_{\bot R}/2 J_R)$. As in the case of staggered dimerization, from (\ref{EpsMF}) and (\ref{EpsCol}) we can establish equivalence between the spectra of the effective Hamiltonian of the columnar ladder and the dimerized $XY$ model in the alternating transverse field:
\begin{eqnarray}
\underline{\mathrm{Columnar~ladder}}~&\Longleftrightarrow&~
\underline{XX ~\mathrm{chain}~\eqref{dgHam},~\gamma=0}\nonumber \\
J_R   ~&\longleftrightarrow&~ J \nonumber \\
\frac12 J_{\bot R} ~&\longleftrightarrow&~ h \nonumber \\
\delta ~&\longleftrightarrow&~ \delta \nonumber
\end{eqnarray}

The columnar phase is trivial not only in the sense that it is always gapped and noncritical, it is
also trivial topologically, since no integer winding number can be assigned to this phase. Indeed, if we identify the opposite ends of the Brillouin zone, then the wavenumber $k$ is defined in the space equivalent to the one-dimensional sphere $S^1$. This is also the space spanned by the two-component unit vector (\ref{n1},\ref{n2}) of the staggered phase. The nontrivial winding number for that phase corresponds to the homotopy group $\pi_1(S^1)=\mathbb{Z}$. As one can see from (\ref{EpsCol}) the analogous unit vector is three-component in the columnar phase, and it spans the space $S^2$. The corresponding homotopy group is trivial $\pi_1(S^2)=0$ and no integer winding number exists, since any closed loop on the sphere $S^2$ can be shrunk.\cite{NakaharaBook}

%
%
\subsection{String Order Parameters}
%
%
%
In order to analytically calculate the SOPs we rewrite the effective Hamiltonian
\eqref{Hmf2} in terms of the Majorana operators
\begin{equation}
\label{Maj}
   a_n +i b_n  \equiv 2 c^{\dag}_n~.
\end{equation}
For the staggered configuration it reads (we drop the constant term $N \mathcal{C}$):
\begin{eqnarray}
\label{HamStMaj}
  H^{\mathrm{st}} &=& \frac{i}{4} J_R(1-\delta)\sum_{l=1}^{N} \big( a_{2l -1} b_{2l+2}+ a_{2l +2} b_{2l-1} \big)
  \nonumber \\
  &-& \frac{i}{4} J_R(1+\delta)\sum_{l=1}^{N} \big( a_{2l} b_{2l+1}+ a_{2l +1} b_{2l} \big)    \nonumber \\
   &-&  \frac{i}{4} J_{\bot R} \sum_{l=1}^{N} \big( a_{2l-1} b_{2l}+ a_{2l} b_{2l-1} \big)~,
\end{eqnarray}
where we labeled the sites of the ladder along the JWT path shown in Fig.~\ref{JWpath}.
Using the JWT \eqref{JWT} in the Majorana representation
\begin{equation}
\label{JWTMaj}
  \left(
    \begin{array}{c}
      \sigma_n^x \\
      \sigma_n^y \\
    \end{array}
  \right)
  =
  \left(
    \begin{array}{c}
      a_n \\
      b_n  \\
    \end{array}
  \right)
  \prod_{l=1}^{n-1} \big[ i a_l b_l \big]
\end{equation}
we can establish the following relations for the inverse JWT from Majorana fermions to the dual spin operators $\tau$
(\ref{sigmaX},\ref{sigmaY}):
\begin{eqnarray}
  \sigma_{n}^{x}  \sigma_{n+1}^{x} &=& i b_n a_{n+1}= \tau_{n-1}^{x}\tau_{n+1}^{x}
  \label{XX} \\
  \sigma_{n}^{y} \sigma_{n+1}^{y} &=& -i a_n b_{n+1} =\tau_{n}^{z}~.
  \label{YY}
\end{eqnarray}
In terms of the dual spin operators, the Hamiltonian \eqref{HamStMaj} maps onto a sum of two
decoupled exactly-solvable Hamiltonians defined on the even and odd sites of the snake-like string
shown in Fig.~\ref{JWpath}:
\begin{widetext}
\begin{eqnarray}
  H^{\mathrm{st}} &=& H^{\mathrm{st}}_{\mathrm{e}} + H^{\mathrm{st}}_{\mathrm{o}}  \nonumber \\
  H^{\mathrm{st}}_{\mathrm{e}}&=&  \frac{1}{4}\sum_{l=1}^{N}
   \big( J_{\bot R}  \tau_{2l-2}^{x} \tau_{2l}^{x}+ J_R (1-\delta)  \tau_{2l-2}^{x} \tau_{2l}^{z} \tau_{2l+2}^{x}
   + J_R (1+\delta) \tau_{2l}^{z}  \big)   \label{HeSt} \\
  H^{\mathrm{st}}_{\mathrm{o}}&=&  \frac{1}{4}\sum_{l=1}^{N}
   \big(  J_R (1+\delta)  \tau_{2l-1}^{x} \tau_{2l+1}^{x} - J_R (1- \delta)  \tau_{2l-1}^{y} \tau_{2l+1}^{y}
       +J_{\bot R} \tau_{2l-1}^{z} \big)~.  \label{HoSt}
\end{eqnarray}
\end{widetext}
The QIM with the three-spin interactions \eqref{HeSt} in the even sector of the Hamiltonian
is discussed in the Appendix. The odd sector \eqref{HoSt} is the $XY$ model in the transverse field with the
anisotropy parameter $\gamma =1 / \delta$.

In our earlier work \cite{GYC11} we took the definition of the SOP for ladders which was proposed by Kim and coworkers\cite{Kim}
and used also in other works.\cite{Delgado} In this paper the ladder SOP is calculated via the string-string correlation function
as defined in Sec.~II. One can check that our SOP is proportional to the SOP of Kim and other workers, but it is given by a simpler
analytical formula. The even and odd SOPs defined below have a clear connection to the even/odd sectors (\ref{HeSt},\ref{HoSt}) of the Hamiltonian.

\begin{figure}[h]
\includegraphics[width=6.0cm]{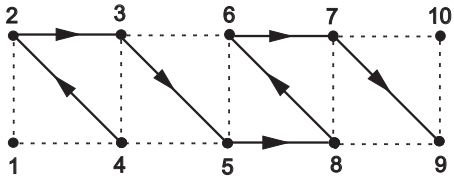}
\caption{String used to define the odd SOP for the two-leg ladder.}
\label{Oddpath}
\end{figure}

To calculate the first SOP we define the string operator \eqref{String} along the path shown in  Fig.~\ref{JWpath}. In so doing we obtain the even SOP:
\begin{equation}
\label{Oeven}
 \mathcal{O}^2_{x,e} =\lim_{ N \rightarrow\infty} (-1)^{N} \Big\{
 \Big\langle \prod_{k=1}^{2N} \sigma_{k}^{x} \Big\rangle =
 \langle \tau^x_{0} \tau^x_{2N}\rangle \Big\}  ~.
\end{equation}
Note that the string chosen allows to incorporate in principle all spins of the ladder in the calculation of $\mathcal{O}_{x,e}$.
Let us now choose another string for the operator \eqref{String} shown in Fig.~\ref{Oddpath}. In such case we obtain the odd SOP:
\begin{equation}
\label{Oodd}
 \mathcal{O}^2_{x,o} =\lim_{ N \rightarrow\infty}  (-1)^{N-1} \Big\{
 \Big\langle \prod_{k=2}^{2N-1} \sigma_{k}^{x} \Big\rangle =
 \langle \tau^x_{1} \tau^x_{2N-1}\rangle \Big\}  ~.
\end{equation}

%
%
%
%

We calculated the order parameter of the three-spin Ising chain \eqref{IsJ3} in the Appendix.
Using the results presented there, we obtain:
\begin{equation}
\label{Oxe}
\mathcal{O}_{x,e} =
\left\{
\begin{array}{lr}
\big[\frac{\kappa^2-\delta^2}{1+ \kappa^2-\delta^2  }  \big]^{1/8}  &  |\kappa / \delta | \geq 1 \\
0 & |\kappa /\delta| <1
\end{array}
\right.
\end{equation}
where $\kappa$ is given by \eqref{kappa}. The odd SOP can be calculated from the results of Barouch and McCoy\cite{Barouch71}
\begin{equation}
\label{Oxo}
\mathcal{O}_{x,o} =
\left\{
\begin{array}{lr}
 0   &  |\kappa / \delta | > 1 \\
  \sqrt{\frac{2}{1+\delta}} \big[\delta^2- \kappa^2 \big]^{1/8}  & |\kappa /\delta| \leq 1
\end{array}
\right.
\end{equation}
In agreement with earlier results \cite{Kim,Delgado,GYC11} we find that
the even and odd SOPs are mutually exclusive. They detect hidden $Z_{2} \otimes Z_{2}$ symmetry
breaking \cite{HiddenSSB} and ordering in the even or odd sectors of the dual Hamiltonian.
Note that contrary to its even counterpart, the string for the odd SOP misses two sites at the
opposite ends of the ladder. The order probed by $\mathcal{O}_{x,o}$ is topologically nontrivial, since it
corresponds to nonzero winding numbers. According to conventional wisdom, \cite{TI} there are zero
Majorana modes at the ladder's ends in this phase. We indicate the order parameters on the phase
diagram of the staggered ladder in Fig.~\ref{PhaseDiagStag}. From the behavior of the gap \eqref{Gap2L}
and SOPs (\ref{Oxe},\ref{Oxo}) near the lines of phase transitions $J_{\bot R}/2J_R = \pm \delta$
we find the critical indices $\nu=1$ and $\beta = 1/8$ of the 2D Ising universality class,
but now their values are the artifact of the effective free-fermionic approximation.

Analysis of the columnar phase is done similarly. The sites of the columnar ladder are labelled
along the JWT path in Fig.~\ref{JWpath}. The Majorana representation of the columnar Hamiltonian \eqref{Hmf2}
can be mapped via transformations (\ref{XX},\ref{YY}) onto dual spins. Again the
dual Hamiltonian splits into even and odd sectors $
H^{\mathrm{col}}= H^{\mathrm{col}}_{\mathrm{o}}+H^{\mathrm{col}}_{\mathrm{e}}$, where the odd term is the
Hamiltonian of the dimerized $XX$ ($\gamma=0$) chain in the alternating field \eqref{dgHam}, while
the even term corresponds to its dual model which is the Ising chain in the alternating transverse
field with the three-spin interactions. Since these models are always gapped (cf. Eq.~\eqref{Gap2LCol}) and do not
undergo any phase  transitions, we will not discuss the properties of the columnar phase in more detail.

Note a  very remarkable property of the effective Hamiltonian \eqref{Hmf2}: for both dimerization patterns it decouples
into the reciprocally dual even and odd sectors
$H^{\mathrm{st/col}}=H^{\mathrm{st/col}}_{\mathrm{o}}+ H^{\mathrm{st/col}}_{\mathrm{e}}$, i.e.
these Hamiltonians  map onto each other
$H^{\mathrm{st/col}}_{\mathrm{o}} \rightleftharpoons H^{\mathrm{st/col}}_{\mathrm{e}}$ under the duality transformation  (\ref{sigmaX},\ref{sigmaY}).

%
%
%
%
\section{Conclusion}\label{Concl}
%
%
In this work we studied quantum phase transitions in the antiferromagnetic dimerized spin-$\frac12$ $XY$ chain and
two-leg ladders. The common feature of these models is a subtle interplay of relevant perturbations each
of which results in a gap creation in the spectrum. There are however lines of quantum criticality where those
perturbations cancel and the models become gapless. We analyzed the properties of the different phases of these
models.

The dimerized $XY$ chain is equivalent to the noninteracting fermions and is mapped by a duality transformation onto
two 1D QIMs residing separately on even and odd sites of the dual lattice. The SOPs for the original chain are mapped onto
the local (Landau) order parameters of these dual QIMs and thus calculated exactly. As follows from the duality mapping,
the dimerized $XY$ model possesses the hidden $\mathbb{Z}_2 \otimes \mathbb{Z}_2$ symmetry, and nonzero SOPs are
due to spontaneous breaking of the $\mathbb{Z}_2$ symmetry in the even or odd sectors of the dual Hamiltonian.

The phase diagram of the model contains topologically trivial phases where the even and odd SOPs coexist along with the conventional long-range order (magnetization). There are
also topologically nontrivial phases without conventional order where only the even or odd nonlocal SOPs exist. In addition
the winding numbers are calculated. Their values are consistent with the topologically trivial or nontrivial nature of each phase.

These analyses and results are quite similar to those for the cluster Ising chain models which attracted attention in recent
literature \cite{Doherty09,Son11,Hamma12,Smacchia11,Lahtinen15} in the context of quantum computation. These exactly solvable models with multi-spin interactions are equivalent to free fermions, and both the conventional and nonlocal SOPs they manifest are readily calculated, similarly to what we have done for the dimerized XY chain.

To extend the framework shown to work so well for the exactly solvable models, the
fermionization of the dimerized two-leg ladders was followed up by a mean-field approximation. The resulting Hamiltonian was then treated as a new effective free-fermionic model. The predictions of this effective Hamiltonian for the critical properties and phase diagram of the ladders with the staggered or columnar dimerizations are in agreement to what is known from earlier work. From the Majorana representation
of the effective Hamiltonian we constructed its transformation to new dual spins. The staggered effective Hamiltonian maps onto the sum of the decoupled even and odd dual exactly-solvable Hamiltonians. The even term is the quantum Ising chain with three-spin interaction, while the odd term is the $XY$ chain in a uniform transverse field. Interestingly, the even and odd sectors map onto each other under the duality transformations. The SOPs of the original spin ladders are calculated as the local order parameters in the dual models.
We defined two mutually exclusive (even and odd) SOPs which detect breaking of the hidden $\mathbb{Z}_2 \otimes \mathbb{Z}_2$ ladder symmetry. One of those SOPs probes topologically trivial phase, while another corresponds to the topologically nontrivial phase. This statement is also corroborated by the calculation of the winding numbers for those phases.  Similar mappings and dualities are found for the columnar dimerized ladder, but this case is not very interesting, since the columnar ladder is always gapped and does not undergo phase transitions.

Our main result is the framework to treat nonlocal orders and hidden symmetries which unifies the key elements
of the Landau paradigm with the new concept of topological order. This unified framework can be
straightforwardly applied for such important physical systems as spin chain and ladders, topological insulators
and superconductors. As we have shown throughout our analysis, the fermionic Hamiltonians (exact or effective) we used for the chain or ladder models are equivalent to tight-binding Hamiltonians of topological materials. To probe the topological order in the latter one needs the Majorana string operators which can be straightforwardly mapped onto the local parameters via the duality transformations we applied in this work.

One more point needs to be emphasized. In the present study we did not need to build up the full-size formalism based on the Ginzburg-Landau effective potential written in terms of the dual spins ($\tau$), since the quadratic Majorana Hamiltonian (exact, as in the case of the $XY$ chain, or an effective mean-field approximation, as in the case of ladders) maps onto exactly solvable dual 1D spin models. So, the key notions of the Landau framework: order parameters, spontaneous symmetry breaking, etc, are easily identified and calculated from those models without any additional approximations. That is why we obtain the critical indices of the 2D Ising universality class (e.g., $\beta =1/8$), but not the mean-field ones ($\beta =1/2$) which a na\"{\i}ve Landau theory would predict.

There are other topologically-ordered lattice systems, including those in higher dimensions which can be mapped via various duality transformations onto the systems with the Landau local orders. \cite{Nussinov09,Nussinov13,Son11}
We are currently working on applying the present approach to the spin-$\frac12$  Heisenberg $n$-leg ladders and tubes. The key step is to obtain the effective mean-field quadratic fermionic Hamiltonian consistent with the $\pi$-flux theorem. \cite{Lieb94}
The transformations to the dual spins maps the Majorana fermion models onto exactly solvable Ising chains with multi-spin interactions.
The local order parameters of these dual models (i.e., the SOPs of the original spin ladders or tubes) can be calculated via Toeplitz determinants. \cite{Lieb61,McCoyBook}
Another interesting developments of the proposed formalism would be its extension for the exotic thermal phases in classical spin models, as, e.g., the algebraically-ordered topological floating phase occurring in some frustrated 2D Ising
models.\cite{BakVillain,UsIsing}

%
%
\begin{acknowledgments}
We thank V. Lahtinen, F. Mila, Z. Nussinov, and J.H.H. Perk for correspondence
and bringing important papers to our attention.
Financial support from NSERC (Canada) and the Laurentian University Research
Fund (LURF) is gratefully acknowledged. T.P. thanks the Ontario Graduate and
the Queen Elizabeth II Fellowships for support.
\end{acknowledgments}
%

\begin {appendix}
\section{Transverse Ising Chain with Three-spin Interactions}\label{App}
%
%
The Hamiltonian of the Ising chain in transverse field with three-spin interactions
is defined as:
\begin{equation}
\label{IsJ3}
   H =  \sum_{i=1}^{N}~ \Big( J \sigma_{i}^{x}\sigma_{i+1}^{x}+
   J_3 \sigma_{i-1}^{x} \sigma_{i}^{z}  \sigma_{i+1}^{x} + h  \sigma_{i}^{z} \Big)~.
\end{equation}
The duality transformations  (\ref{sigmaX},\ref{sigmaY} map the above Hamiltonian
onto the anisotropic $XY$ chain in transverse field. \cite{Peschel04} Some basic  properties of the
model  \eqref{IsJ3} were analyzed in \cite{Kopp05}  and later in \cite{Dutta07,Niu12}.
The Jordan-Wigner transformation of \eqref{IsJ3} yields the free-fermionic
model (up to an unimportant constant term):
\begin{equation}
\label{IsJ3JW}
   H =  \sum_{q}~ \Big( 2 A(q)c_q^\dag c_q- i B(q)(c_q^\dag c_{-q}^\dag+c_q c_{-q}) \Big)~,
\end{equation}
where
\begin{eqnarray}
  A(q) &\equiv& h+J \cos q-J_3  \cos 2q
  \label{A} \\
  B(q) &\equiv& J \sin q -J_3  \sin 2q~,
  \label{B}
\end{eqnarray}
and $c_q^\dag,c_q$ are the Fourier transforms of the lattice fermion operators $c_i^\dag,c_i$.
The Bogoliubov transformation
\begin{equation}
\label{Bogol}
   \eta_q = \cos \Theta_q c_q -i \sin \Theta_q c_{-q}^\dag
\end{equation}
with the  Bogoliubov angle $\Theta_q$ defined as $\tan 2\Theta_q =B(q)/A(q)$, diagonalizes the Hamiltonian. The
spectrum of the quasiparticles $\eta_q, \eta_q^\dag$ is
\begin{eqnarray}
   \varepsilon(q) = \sqrt{A(q)^2+B(q)^2} =\nonumber \\
  \sqrt{h^2+J^2+J_3^2+2J(h-J_3)\cos q- 2h J_3 \cos 2q}
  \label{Eps3}
\end{eqnarray}
The model is gapless along the lines of quantum criticality $h=J_3 \pm J$, and for the
transverse fields within the range
\begin{equation}
\label{Order}
   J_3-J <h <J_3+J
\end{equation}
the model manifests the local long-range order\cite{Kopp05} (magnetization) $m_x \neq 0$.
Our goal is to calculate the order parameter $m_x$.  We will use the classical approaches.
\cite{Lieb61,Wu66,Barouch71} More modern treatments and references can be found, e.g.,
in \cite{McCoyBook,Abanov05}. The fermionic correlation function
\begin{equation}
\label{Glm}
   G_{l-m} \equiv i \langle b_l a_m\rangle
\end{equation}
is defined in terms of the site Majorana operators \eqref{Maj}.
It can be calculated as the Fourier transform
\begin{equation}
\label{Gn}
   G_n = \int_{0}^{2 \pi}  \frac{d \varphi}{2 \pi} e^{i n \varphi} G(\varphi)
\end{equation}
of the generating function
\begin{equation}
\label{Gphi}
   G(\varphi) = \sqrt{\frac{A(\varphi)+iB(\varphi)}{A(\varphi)-iB(\varphi)}}
   = \Bigg[\frac{(e^{i \varphi}-\alpha_+)(e^{i \varphi}-\alpha_-)
   }{(e^{-i \varphi}-\alpha_+)(e^{-i \varphi}-\alpha_-)}  \Bigg]^{1/2}
\end{equation}
where
\begin{equation}
\label{alphapm}
   \alpha_{\pm} \equiv  \frac{J \pm \sqrt{J^2+4hJ_3}}{2J_3}~.
\end{equation}
The spin-spin correlation function $\langle \sigma_1^x \sigma_{N+1}^x \rangle$
is given by the $N \times N$ Toeplitz determinant:\cite{Lieb61}
\begin{equation}
\label{sigmaTep}
   \langle \sigma_1^x \sigma_{N+1}^x \rangle= \left|
   \begin{array}{ccc}
     G_{-1} & ... & G_{-N} \\
     : & ... & : \\[-0.30cm]
     . & ~ & . \\
     G_{N-2} & ... & G_{-1}
   \end{array}
   \right| \equiv D_N
   ~.
\end{equation}
Note that the singular points of $G(\varphi)$ satisfy
$\alpha_+ \alpha_- =-h/J_3$, and within the magnetically ordered range \eqref{Order}
the first root is bound $-1 <\alpha_- <0$, thus for $J_3/h <1$ the inequality
$0 <\alpha_+^{-1} <1$ for the second root holds. The generating function can be written
as $G(\varphi)= e^{i \varphi} \tilde{G}(\varphi)$ with
\begin{equation}
\label{GphiTilda}
   \tilde{G}(\varphi) =
   \Bigg[\frac{(e^{- i \varphi}-\alpha_+^{-1})(e^{i \varphi}-\alpha_-)
   }{(e^{ \varphi}-\alpha_+^{-1})(e^{-i \varphi}-\alpha_-)}  \Bigg]^{1/2}~.
\end{equation}
Then
\begin{equation}
\label{GnTilda}
   G_n = \int_{0}^{2 \pi}  \frac{d \varphi}{2 \pi} e^{i (n+1) \varphi} \tilde{G}(\varphi)=\tilde{G}_{n+1}~,
\end{equation}
and the determinant \eqref{sigmaTep} can be written in terms of $\tilde{G}_n$:
\begin{equation}
\label{DTilda}
   D_N
   = \left|
   \begin{array}{ccc}
     \tilde{G}_{0} & ... & \tilde{G}_{-N+1} \\
     : & ... & : \\[-0.30cm]
     . & ~ & . \\
     \tilde{G}_{N-1} & ... & \tilde{G}_{0}
   \end{array}
   \right| \equiv \tilde{D}_N
   ~.
\end{equation}
Since zeros/singularities of the generating function  $\tilde{G}(\varphi)$ $\{|\alpha_+^{-1}|,|\alpha_-| \}<1$ lie inside
the unit circle $z=e^{i \varphi}$ on the complex plane, the conditions for Szeg\"{o}'s theorem are satisfied \cite{McCoyBook},
and it can be applied for calculation of the limit $N \to \infty$ of the Toeplitz determinant $\tilde{D}_N$.
The latter result is well known and reads:\cite{McCoyBook}
\begin{equation}
\label{M2}
   \lim_{N \to \infty} \tilde{D}_N =  \Bigg[\frac{(1- \alpha_+^{-2})(1-\alpha_-^2)
   }{(1-\alpha_-/\alpha_+)^2}  \Bigg]^{1/4}~.
\end{equation}
Combining all formulae together, we obtain the sought result:
\begin{equation}
\label{mx}
   m_x^2= \lim_{N \to \infty}  \langle \sigma_1^x \sigma_{N+1}^x \rangle=
  \Bigg[\frac{J^2-(J_3-h)^2}{J^2+4 h J_3}  \Bigg]^{1/4}~.
\end{equation}
The above formula for magnetization is analytic in the whole locally ordered range \eqref{Order} across
the line $h=J_3$. The validity of the formula for the order parameter \eqref{mx} was checked by direct numerical
calculation of the determinant \eqref{sigmaTep} using \textsl{Mathematica}.
\end{appendix}

%

%
%

\begin{thebibliography}{}

%
\bibitem{Landau5} L.D. Landau and E.M. Lifshitz, \textit{Statistical
Physics Part 1.} Course of Theoretical Physics Vol.~5 (3rd ed.),
Butterworth-Heinemann (1980).
%
%
\bibitem{Wen}
X.-G.Wen, \textit{Quantum Field Theory of Many-Body Systems} (Oxford, New York,
2004); P.A. Lee, N. Nagaosa, and X.-G. Wen, Rev. Mod. Phys. \textbf{78}, 17
(2006).
%
%
\bibitem{FradkinBook13} E. Fradkin, \textit{Field Theories of Condensed Matter Physics}, 2nd
edition (Cambridge University Press, New York, 2013).
%

%
\bibitem{Dagotto} E. Dagotto and T.M. Rice, Science \textbf{271},
618 (1996); E. Dagotto, Rep. Prog. Phys. \textbf{62}, 1525 (1999).
%
%
\bibitem{Laflorencie2011}
A. Lavar\'elo, G. Roux, and N. Laflorencie, Phys. Rev. B \textbf{84}, 144407
(2011).
%
\bibitem{LaddersQC} Y.-J. Wang, F. H. L. Essler, M. Fabrizio, and A. A. Nersesyan,
Phys. Rev. B \textbf{66}, 024412 (2002); V. Gritsev, B. Normand, and D.
Baeriswyl, Phys. Rev. B \textbf{69}, 094431 (2004); O.A. Starykh and L.Balents,
Phys. Rev. Lett. \textbf{93}, 127202 (2004); M. Sato, Phys. Rev. B \textbf{76},
054427 (2007); G.-H. Liu, H.-L. Wang, and G.-S. Tian, Phys. Rev. B \textbf{77},
214418 (2008); T. Hikihara and O.A. Starykh, Phys. Rev. B \textbf{81}, 064432
(2010).
%
%
\bibitem{SOPladders} H. Watanabe, Phys. Rev. B {\bf52}, 12508 (1995);
Y. Nishiyama, N. Hatano and M. Suzuki, J. Phys. Soc. Jpn. {\bf64}, 1967 (1995);
D. G. Shelton, A. A. Nersesyan, and A. M. Tsvelik, Phys. Rev. B \textbf{53},
8521 (1996).
%
%
\bibitem{Kim}
E.H. Kim, G. Fath, J. Solyom, and D. J. Scalapino, Phys. Rev. B \textbf{62},
14965 (2000); G. Fath, O. Legeza, and J. Solyom, Phys. Rev. B {\bf63}, 134403
(2001);  E.H. Kim, O. Legeza, and J. Solyom, Phys. Rev. B \textbf{77}, 205121
(2008).
%
%
\bibitem{Delgado} M.A. Martin-Delgado, R. Shankar, and G. Sierra,
Phys. Rev. Lett. \textbf{77}, 3443 (1996); M.A. Martin-Delgado, J. Dukelsky,
and G. Sierra, Phys. Lett. A \textbf{250}, 430 (1998); J. Almeida, M.A.
Martin-Delgado, and G. Sierra, Phys. Rev. B \textbf{76}, 184428 (2007);
\textit{ibid} \textbf{77}, 094415 (2008); J. Phys. A \textbf{41}, 485301
(2008).
%
%
\bibitem{2Dspins}
G. Misguich and C. Lhuillier, in \textit{Frustrated Spin Systems}, edited by
H.T. Diep (World Scientific, Singapore, 2005), p. 229.
%
%
\bibitem{Kitaev06}
A. Kitaev, Ann. Phys. \textbf{321}, 2 (2006).
%
%
\bibitem{Xiang07} X.-Y. Feng, G.-M. Zhang, and T. Xiang,
Phys. Rev. Lett. \textbf{98}, 087204 (2007).
%
%
\bibitem{TI} For reviews and refs:
X.-L. Qi and S.-C. Zhang, Rev. Mod. Phys. \textbf{83}, 1057 (2011); M.Z. Hasan
and C.L. Kane, \textit{ibid}, \textbf{82} 3045 (2010); B.A. Bernevig and T.L.
Hughes, \textit{Topological Insulators and Topological Superconductors}
(Princeton University Press, Princeton, 2013); M. Fruchart and D. Carpentier,
Comptes Rendus Physique \textbf{14}, 779 (2013).
%
%
\bibitem{Nussinov09}
Z. Nussinov and G. Ortiz, Proc. Natl. Acad. Sci. USA \textbf{106}, 16944 (2009);
Ann. of Phys. \textbf{324}, 977 (2009).
%
%
\bibitem{VolovikBook}  G.E. Volovik, \textit{The Universe in a Helium Droplet}
(Clarendon Press, Oxford, 2003).
%
%
\bibitem{Hatsugai-1} Y. Hatsugai, J. Phys. Soc. \textbf{75}, 123601 (2006);
T. Hirano, H. Katsura, and Y. Hatsugai, Phys. Rev. B \textbf{77}, 094431
(2008).
%
%
\bibitem{Hatsugai-2}
M. Arikawa, I. Maruyama, and Y. Hatsugai, Phys. Rev.  B \textbf{82},
073105 (2010); I. Maruyama, T. Hirano, and Y. Hatsugai, Phys. Rev.  B
\textbf{79}, 115107 (2009).
%
%
\bibitem{Ezawa13}
M. Ezawa, Y. Tanaka, and N. Nagaosa, Sci. Rep. \textbf{3}, 2790 (2013).
%
%
\bibitem{Mila13} N. Chepiga, F. Michaud, and F. Mila,
Phys. Rev. B \textbf{88}, 184418 (2013).
%
%
\bibitem{Entang} For reviews: L. Amico, \textit{et al}, Rev. Mod. Phys. \textbf{80}, 517 (2008);
J.I. Latorre and A. Riera, J. Phys. A \textbf{42}, 504002 (2009).
%
%
\bibitem{denNijs89}
M. den Nijs and K. Rommelse, Phys. Rev. B \textbf{40}, 4709 (1989).
%
%
\bibitem{HiddenSSB} M. Oshikawa, J. Phys. Condens. Matt. \textbf{4}, 7469
(1992); T. Kennedy and H. Tasaki, Phys. Rev. B \textbf{45}, 304 (1992); M.
Kohmoto and H. Tasaki, Phys. Rev. B \textbf{46}, 3486 (1992).
%
%
\bibitem{Hida92} K. Hida, Phys. Rev. B \textbf{45}, 2207 (1992).
%
%
\bibitem{GYC11}
S.J. Gibson, R. Meyer, and G.Y. Chitov, Phys. Rev. B \textbf{83}, 104423
(2011).
%
%
\bibitem{Bortz07} M. Bortz, J. Sato, and M. Shiroishi,
J. Phys. A  \textbf{40}, 4253 (2007).
%
%
\bibitem{Kotov99} V.N. Kotov, J. Oitmaa, and Z. Weihong, Phys. Rev. B
\textbf{59}, 11377 (1999).
%
%
\bibitem{Cabra99} D.C. Cabra and M.D. Grynberg,
Phys. Rev. Lett. \textbf{82}, 1768 (1999).
%
%
\bibitem{Nersesyan00} Y.-J. Wang and A.A. Nersesyan, Nucl. Phys. B
\textbf{583} [FS], 671 (2000).
%
%
\bibitem{Okamoto03} K. Okamoto, Phys. Rev. B \textbf{67}, 212408
(2003).
%
%
\bibitem{Nakamura03} M. Nakamura, T. Yamamoto, and K. Ide,
J. Phys. Soc. Jpn.  \textbf{72}, 1022 (2003).
%
%
\bibitem{GYC08}
G.Y. Chitov, B.W. Ramakko, and M. Azzouz, Phys. Rev. B \textbf{77}, 224433
(2008).
%
%
\bibitem{MoreDimLadd}
J. Chen, K.-L. Yao, and L.-J. Ding, Physica A \textbf{391}, 2306 (2012);
Y.J. Xu, H. Zhao, Y.G. Chen, and Y.H. Yan. Eur. Phys. J. B \textbf{85}, 393 (2012);
R.-X. Li, S.-L. Wang, K.-L. Yao, and H.-H. Fu, Phys. Lett. A \textbf{377}, 2422 (2013).
%
%
\bibitem{Berg08} E. Berg, E.G. Dalla Torre, T. Giamarchi, and E. Altman,
Phys. Rev. B \textbf{77}, 245119 (2008).
%
%
\bibitem{Rath13} S.P. Rath, W. Simeth, M. Endres, and
W. Zwerger, Annals of Physics \textbf{334}, 256 (2013).
%
%
\bibitem{Enders11}
M. Endres, \textit{et al}, Science \textbf{334}, 200 (2011).
%
%
\bibitem{UsPerc}
P.N. Timonin and G.Y. Chitov, J. Phys. A: Math. Theor. \textbf{48}, 135003 (2015); Phys. Rev. E \textbf{93}, 012102 (2016).
%
%
\bibitem{ChenHu07}
H.-D. Chen and J. Hu, Phys. Rev. B \textbf{76}, 193101 (2007).
%
%
\bibitem{GYC07} M. Azzouz, K. Shahin, and G.Y. Chitov, Phys. Rev. B
\textbf{76}, 132410 (2007).
%
%
\bibitem{Perk75} J.H.H. Perk, H.W. Capel, M.J. Zuilhof, and Th. J. Siskens,
Physica A \textbf{81}, 319 (1975).
%
%
\bibitem{DelGamMod}
F. Ye, G.-H. Ding, and B.-W. Xu, Commun. Theor. Phys. (Beijing, China)
\textbf{37}, 492 (2002); F. Ye and B.-W. Xu, Commun. Theor. Phys. (Beijing,
China) \textbf{39}, 487 (2003).
%
%
\bibitem{Lieb61} E.H.  Lieb, T. Schultz, and D. Mattis, Ann. Phys. (N.Y.) \textbf{16}, 407 (1961).
%
%
\bibitem{Fradkin78} E. Fradkin, and L. Susskind, Phy. Rev. D \textbf{17}, 2637
(1978).
%
%
\bibitem{Pfeuty70} P. Pfeuty, Ann. Phys. (N.Y.) \textbf{ 57}, 79 (1970).
%
%
\bibitem{Barouch71} E. Barouch  and B.M. McCoy, Phys. Rev. A \textbf{3}, 786 (1971).
%
%
\bibitem{Wu12} N. Wu, Phys. Lett. A \textbf{376}, 3530 (2012).
%
%
\bibitem{Niu12}
Y. Niu, S. B. Chung, C.-H. Hsu, I. Mandal, S. Raghu, and S. Chakravarty, Phys.
Rev. B \textbf{85}, 035110 (2012).
%
%
\bibitem{SchnyderRyu11}
A.P. Schnyder and S. Ryu, Phys. Rev. B \textbf{84}, 060504(R) (2011).
%
\bibitem{Abanov05} F. Franchini and A.G. Abanov, J. Phys. A \textbf{38}, 5069 (2005).
%
%
\bibitem{Azzouz}
M. Azzouz, Phys. Rev. B \textbf{48}, 6136 (1993); M. Azzouz, L. Chen, and S.
Moukouri, Phys. Rev. B \textbf{50}, 6233 (1994).
%
%
\bibitem{Dai98Etal} X. Dai and Z. Su, Rev. B \textbf{57}, 964 (1998);
H. Hori and S. Yamamoto, J. Phys. Soc. \textbf{73}, 549 (2004); T.S. Nunner and
T. Kopp, Rev. B \textbf{69}, 104419 (2004).
%
%
\bibitem{Lieb94} E.H. Lieb, Phys. Rev. Lett. \textbf{73}, 2158 (1994).
%
%
\bibitem{Affleck88} I. Affleck and J.B. Marston, Phys.Rev. B \textbf{37}, 3774 (1988).
%
%
\bibitem{Toplal14} T. Pandey, M.S. thesis, Laurentian University, 2014.
%
%
\bibitem{NakaharaBook} M. Nakahara, \textit{Geometry, Topology and Physics},
2nd edition (Taylor \& Francis, New York, 2003).
%
%
\bibitem{Doherty09}
A.C. Doherty and S.D. Bartlett, Phys. Rev. Lett. \textbf{103}, 020506 (2009).
%
%
\bibitem{Son11}
W. Son, L. Amico, R. Fazio, A. Hamma, S. Pascazio, and V. Vedral, Europhys. Lett.
 \textbf{95}, 50001 (2011).
%
%
\bibitem{Hamma12}
S. Montes and A. Hamma, Phys. Rev. E \textbf{86}, 021101 (2012).
%
%
\bibitem{Smacchia11}
P. Smacchia, L. Amico, P. Facchi, R. Fazio, G. Florio, S. Pascazio, and V. Vedral,
Phys. Rev. A \textbf{84}, 022304 (2011).
%
%
\bibitem{Lahtinen15}
V. Lahtinen and E. Ardonne, Phys. Rev. Lett. \textbf{115}, 237203 (2015).
%
%
\bibitem{Nussinov13}
E. Cobanera, G. Ortiz, and Z. Nussinov, Phys. Rev. B \textbf{87}, 041105(R) (2013).
%
%
\bibitem{BakVillain} J. Villain and P. Bak, J. de Phys.\textbf{ 42}, 657
(1981); P. Bak, Rep. Progr. Phys. \textbf{45}, 587 (1982).
%
%
\bibitem{UsIsing}
A. Kalz and G.Y. Chitov, Phys. Rev. B \textbf{88}, 014415 (2013);
G.Y. Chitov and C. Gros, Low Temperature Physics {\bf 31}, 722 (2005).
%
%
\bibitem{Peschel04} I. Peschel, J. Stat. Mech.: Theor. Exp. P12005 (2004).
%
%
\bibitem{Kopp05} A. Kopp and S. Chakravarty, Nature Phys. \textbf{1}, 53 (2005).
%
%
\bibitem{Dutta07} U. Divakaran and A. Dutta, J. Stat. Mech.: Theor. Exp. P11001 (2007).
%
%
\bibitem{Wu66} T.T. Wu, Phys. Rev. \textbf{149}, 380 (1966).
%
%
\bibitem{McCoyBook} B.M. McCoy, \textit{Advanced Statistical Mechanics}
(Oxford University Press, New York, 2010).
%
%
%
%






\end{thebibliography}
\end{document}